\pdfoutput=1

\documentclass[11pt]{article}

\usepackage{EACL2023}

\usepackage{times}
\usepackage{latexsym}
\usepackage{graphicx} 

\usepackage[T1]{fontenc}

\usepackage[utf8]{inputenc}

\usepackage{microtype}

\usepackage{inconsolata}

\usepackage{multirow}
\usepackage{forloop}
\usepackage{enumitem}

\title{What Makes A Video Radicalizing? Identifying Sources of Influence in QAnon Videos}

\author{
Lin Ai\textsuperscript{1},
Yu-Wen Chen\textsuperscript{1},
Yuwen Yu\textsuperscript{2},
Seoyoung Kweon\textsuperscript{1},
\\
\textbf{Julia Hirschberg\textsuperscript{1},}
\textbf{Sarah Ita Levitan\textsuperscript{2}}
\\
\textsuperscript{1}Columbia University,
\textsuperscript{2}Hunter College
\\
\{lin.ai, julia\}@cs.columbia.edu, \{yc4093, sk4865\}@columbia.edu
\\
yyu4@gradcenter.cuny.edu, sarah.levitan@hunter.cuny.edu
}

\begin{document}
\maketitle
\begin{abstract}

In recent years, radicalization is being increasingly attempted on video-sharing platforms. Previous studies have been proposed to identify online radicalization using generic social context analysis, without taking into account comprehensive viewer traits and how those can affect viewers' perception of radicalizing content. To address the challenge, we examine QAnon, a conspiracy-based radicalizing group, and have designed a comprehensive questionnaire aiming to understand viewers' perceptions of QAnon videos. We outline the traits of viewers that QAnon videos are the most appealing to, and identify influential factors that impact viewers' perception of the videos. 

\end{abstract}

\section{Introduction}



Radicalization, the process of developing extremist ideologies and beliefs in others, has been increasingly seen on social media in recent years. Previous studies have proposed to identify online radicalization using lexical and social context analysis. However, much of the current radicalization is being attempted on video-sharing platforms, where multimodality features beyond text can be powerful in the promotion of extremist content. Moreover, generic social context analysis does not take into account comprehensive viewer traits and how those can affect viewers' perception of radicalizing content. To address these challenges, we focus on radicalization in YouTube and BitChute. We examine QAnon, a conspiracy-based radicalizing group originated in 2017. We have collected a QAnon video corpus from YouTube and BitChute, and have designed a comprehensive questionnaire aiming to identify traits of viewers that QAnon videos are the most appealing to, influential factors that contribute to viewers' perception, and how these traits differ between pro- and anti-QAnon videos. To the best of our knowledge, this is the first work aiming to computationally analyze viewers' perception of QAnon video.

In this study, we focus on three main research questions: \textbf{RQ1:} What viewer traits, such as personality traits and media consumption, are associated with their video preferences? \textbf{RQ2:} What video characteristics, such as speaker traits, video quality, and arousing emotions, are correlated with viewers' perception? \textbf{RQ3:} Which modality features affect viewers' perception the most?

\section{Related Work}
Much work has been done on radicalization in social media. \citet{hartung-etal-2017-ranking} attempt to identify right-wing extremist content in German Twitter profiles; \citet{hofmann-etal-2022-modeling} leverage network structure of Reddit forums to detect polarized concepts; and \citet{lopez2018towards} and \citet{araque2020approach} develop methods to identify radicalizing content in Twitter. Research has also been done using multimodal features to detect radicalization in Jihadist YouTube videos using social network analysis and sentiment \cite{bermingham2009combining}. \citet{ribeiro2020auditing} collect 330,925 YouTube videos to identify radicalizing pipelines for far-right groups, and \citet{ai2021identifying} identify multimodal features of far-right and far-left groups which them more popular and more persuasive.  

In recent years, QAnon has been identified as one of the prime conspiracy-based radicalization groups \cite{amarasingam2020qanon, garry2021qanon}. However, little study has computationally analyzed QAnon related videos, in terms of how these videos drag viewers into the process of radicalization, and who the videos are the most appealing to. Therefore, in this work, we aim to identify the viewers that are attracted the most to QAnon videos, and influential factors of the videos that contribute the most to viewers' perception.








\section{Corpus and Annotation Collection}
We have collected 5,924 YouTube and BitChute videos on QAnon to study a full range of multimodal characteristics of QAnon videos. We then select a small subset of these videos, 3 pro- and 3 anti-QAnon, based on the videos' relevance to the topic, duration, diversity in styles, quality of content, and popularity measured by number of likes, comments and shares.  To obtain human rating, we create a comprehensive questionnaire asking raters to explain aspects of their perception of the videos and of QAnon, and the actions they believe that they or others might take after watching the videos. The questionnaire is included in Appendix \ref{appendix:survey-questions}.

\subsection{Rater Demographics and Background}
\label{subsection:rater-demographics}

A total of 46 raters take part in the questionnaire. In the beginning of the questionnaire, we ask raters a few questions about their own demographics, including gender, age, ethnicity, level of education, and political leaning. See Table \ref{tab:survey-introductory} in Appendix \ref{appendix:survey-questions} for the full question set. The distribution of rater demographics is shown in Figure \ref{fig:demographics} in Appendix \ref{appendix:rater-demographics}.
We also ask raters to provide personality information, as we are interested in learning a comprehensive profile of viewers that would be attracted to either pro- or anti-QAnon videos. For this, raters complete the Ten Item Personality Inventory \cite{gosling2003very}, that measures the Big Five personality dimensions: Neuroticism, Extraversion, Openness to Experience, Agreeableness, and Conscientiousness. The responses are summarized in Figure \ref{fig:self-report-personality} in Appendix \ref{appendix:rater-demographics}. 

\begin{figure*}[!htb]
\centering
\includegraphics[width=.8\textwidth]{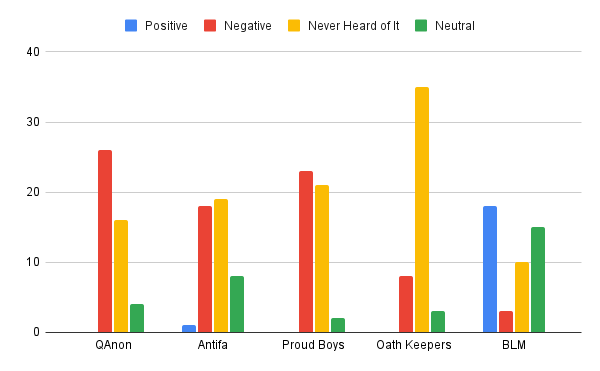}\hfill
\caption{Rater's opinion on radical groups. A total of 46 raters completed the questionnaire.}
\label{fig:radical-group-opinion}
\end{figure*}

\begin{figure*}[!htb]
\centering
\includegraphics[width=.8\textwidth]{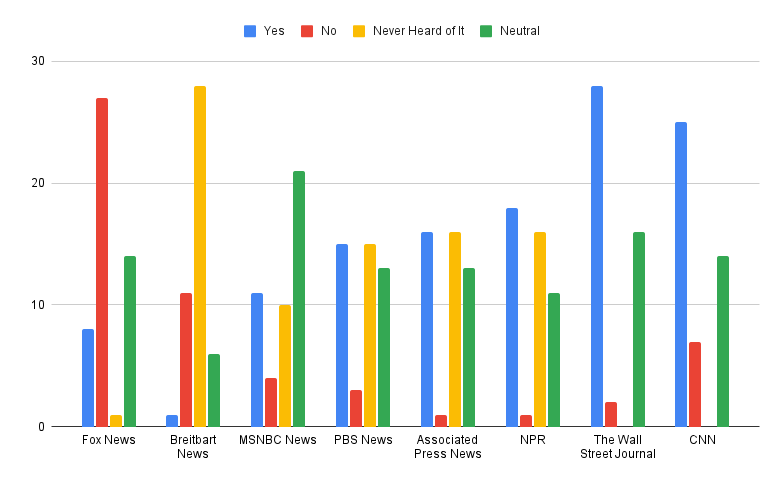}\hfill
\caption{Rater's opinion on media sources. A total of 46 raters completed the questionnaire.}
\label{fig:media-opinion}
\end{figure*}

To study how individuals' perception of potentially radical videos may be affected by their initial impression of extremist groups and the media they consume, we ask raters to rate their opinions, positive, negative or neutral, of five well-known extremist groups, and how much they trust eight of the mainstream media sources. The five extremist groups include three far-right groups (QAnon, The Proud Boys, Oath keepers) and two far-left groups (Antifa, the subset of BLM that involves in violent actions); and the eight media sources are Fox News, Breitbart News, MSNBC News, PBS News, Associated Press News (AP), NPR, The Wall Street Journal (WSJ), and CNN. The political bias of these media sources are obtained from Media Bias/Fact Check (MBFC)\footnote{\href{https://mediabiasfactcheck.com/}{https://mediabiasfactcheck.com/}}. The responses are summarized in Figure \ref{fig:radical-group-opinion} and \ref{fig:media-opinion}.

\subsection{Evaluation Metrics}
As \citet{borum2011radicalization} argues, radicalization needs to be distinguished from action pathways, the process of engaging in violent extremist actions, as most people with radical ideas do not engage in violent actions or terrorism. Being curious about certain extremist groups, or even considering joining the groups, are often the first steps in such action pathways. In this study, we generalize the concept of radicalization as the process of developing extremist ideologies and taking the first steps in the action pathways towards violence. Therefore, to better assess the level of radicalization of a video, we separately evaluate viewers' overall impression towards the video, including whether they enjoy watching the video in general, how they feel about the content of the video, and the actions they think they would take after watching the video. With this purpose, we use 3 metrics:
\begin{enumerate} 
    \item \textbf{Enjoyment Score:} raters are asked to rate how much they enjoy watching each video on a 5-point Likert scale. The Enjoyment Scores are converted to [-2, 2].
    \item \textbf{Content Score:} raters are asked to say whether they think a video is persuasive, trustworthy, logical, and professionally created and these rating scores are each converted to [-1, 1]. Each video's Content Score is the sum of these 4 traits' scores. High Content Scores imply that raters agree with the video content and think that it was was valid, trustworthy, persuasive, and logical. 
    \item \textbf{Actions Score:} raters are asked whether they would take the following actions after watching a video, listed from the most active gourp opposing actions to the most active group supporting actions: \textbf{a)} posting a criticizing comment [score -2] \textbf{b)} disliking the video [score -1] \textbf{c)} liking the video [score 1] \textbf{d)} posting a supporting comment [score 2] \textbf{e)} considering joining the group [score 3]. The Actions Score of a video is the sum of these actions' scores. The higher the Actions Score, the more actively the raters support the video, or even the QAnon ideology.
\end{enumerate}

\section{Analyzing Viewer Ratings and Traits}
In this and the following sections, we use the words rater and viewer interchangeably. To answer \textbf{RQ1}, we investigate how viewers' self-reported personalities, initial impression of extremist groups, and their media consumption correlate with their preference for QAnon videos. We examine how these traits correlate with the Enjoyment Scores, Content Scores, and Actions Scores they give to all QAnon videos as well as just to pro- or anti-QAnon videos. For each metric score, we calculate a viewers' overall score on all videos, pro-QAnon videos, and anti-QAnon videos as the average score they give to each video, to each pro-QAnon video, and to each anti-QAnon video. 

We perform significance tests on the Spearman's correlation between these viewer traits and the three metric scores. For our Enjoyment Score, the significant viewer traits (p-value < 0.05) are presented in Table \ref{tab:sig-test-viewer-enjoy}. Viewers having a positive opinion of Antifa and of The Proud Boys enjoy watching all our QAnon videos in general. Particularly, viewers with a positive opinion towards Antifa enjoy watching anti-QAnon videos. 
This matches our impression because The Proud Boys is also a far-right group, thus, viewers supporting The Proud Boys enjoy watching QAnon videos in general; whereas Antifa is a left-wing group, thus, viewers supporting Antifa enjoy watching anti-QAnon videos. Viewers trusting CNN news tend to enjoy watching QAnon videos, especially, the pro-QAnon videos, which is somewhat surprising since CNN is a left-biased media. One possible explanation could be that sometimes, people might feel hilarious when perceiving information from the opposite side. Other viewers enjoying watching pro-QAnon videos are those who trust the WSJ, aligning with our assumption that right-leaning viewers would trust a right-center based source. 

\begin{table}[!htb]
\small
\centering
\begin{tabular}{ccc}
\hline
\multicolumn{3}{c}{\textbf{Enjoyment on All Videos}}  \\
\textbf{Feature}   & \textbf{Corr} & \textbf{p-value} \\
\hline
Opinion\_CNN       & 0.358      & 0.0146         \\
Opinion\_Antifa    & 0.345      & 0.0189         \\
Opinion\_ProudBoys & 0.297      & 0.0452         \\
\hline\hline
\multicolumn{3}{c}{\textbf{Enjoyment on Pro-QAnon Videos}}  \\
\textbf{Feature}   & \textbf{Corr} & \textbf{p-value} \\
\hline
Opinion\_CNN       & 0.329      & 0.0255         \\
Opinion\_WSJ       & 0.298      & 0.0440         \\
\hline\hline
\multicolumn{3}{c}{\textbf{Enjoyment on Anti-QAnon Videos}} \\
\textbf{Feature}   & \textbf{Corr} & \textbf{p-value} \\
\hline
Opinion\_Antifa    & 0.368      & 0.0119         \\ 
\hline
\end{tabular}
\caption{Significant viewer ratings and traits (p-value < 0.05) on Enjoyment Scores}
\label{tab:sig-test-viewer-enjoy}
\vspace{-0.5mm}
\end{table}

For our Content Score, the significant viewer ratings and traits are listed in Table \ref{tab:sig-test-viewer-content}. Generally, viewers who trust Fox News agree with the content of our selected QAnon videos, specifically, pro-QAnon videos. This agrees with our presumption, as Fox News is rated as right-biased media. On the other hand, viewers trusting NPR and AP tend to disagree with the content of pro-QAnon videos, which makes sense, since both media sources are left-center biased. In addition, viewers who are self-reported as reserved and quiet tend to agree with the content of anti-QAnon videos.

\begin{table}[!htb]
\small
\centering
\begin{tabular}{ccc}
\hline
\multicolumn{3}{c}{\textbf{Content of All Videos}}  \\
\textbf{Feature} & \textbf{Corr} & \textbf{p-value} \\
\hline
Opinion\_Fox     & 0.430      & 0.00283         \\
\hline\hline
\multicolumn{3}{c}{\textbf{Content of Pro-QAnon Videos}}  \\
\textbf{Feature} & \textbf{Corr} & \textbf{p-value} \\
\hline
Opinion\_Fox     & 0.487      & 0.000592         \\
Opinion\_NPR     & -0.376     & 0.0100           \\
Opinion\_AP      & -0.330     & 0.0253           \\
\hline\hline
\multicolumn{3}{c}{\textbf{Content of Anti-QAnon Videos}} \\
\textbf{Feature} & \textbf{Corr} & \textbf{p-value} \\
\hline
Reserved         & 0.339      & 0.0213         \\
\hline
\end{tabular}
\caption{Significant viewer traits and ratings (p-value < 0.05) on Content Scores}
\label{tab:sig-test-viewer-content}
\vspace{-0.5mm}
\end{table}

For our Actions Score, the significant viewer ratings and traits are listed in Table \ref{tab:sig-test-viewer-actions}. As we expect, viewers with positive opinions towards Oath Keepers, Fox News, and WSJ tend to actively support selected QAnon videos, especially pro-QAnon videos, because Oath Keepers is considered a far-right group, and Fox News and WSJ are both right leaning. Surprisingly, viewers with positive opinions towards Antifa and CNN also tend to support pro-QAnon videos. In addition, viewers self-reported as disorganized and careless tend to support anti-QAnon videos, and viewers self-reported as sympathetic and warm tend to oppose anti-QAnon videos.

\begin{table}[!htb]
\small
\centering
\begin{tabular}{ccc}
\hline
\multicolumn{3}{c}{\textbf{Actions after All Videos}}   \\
\textbf{Feature}     & \textbf{Corr} & \textbf{p-value} \\
\hline
Opinion\_OathKeepers & 0.387      & 0.00793         \\
Opinion\_Antifa      & 0.359      & 0.0143          \\
Opinion\_Fox         & 0.350      & 0.0172          \\
Opinion\_WSJ         & 0.322      & 0.0291          \\
\hline\hline
\multicolumn{3}{c}{\textbf{Actions after Pro-QAnon Videos}}   \\
\textbf{Feature}     & \textbf{Corr} & \textbf{p-value} \\
\hline
Opinion\_OathKeepers & 0.370      & 0.0114         \\
Opinion\_Fox         & 0.358      & 0.0145         \\
Opinion\_WSJ         & 0.346      & 0.0186         \\
Opinion\_CNN         & 0.298      & 0.0442         \\
Opinion\_Antifa      & 0.295      & 0.0467         \\
\hline\hline
\multicolumn{3}{c}{\textbf{Actions after Anti-QAnon Videos}}  \\
\textbf{Feature}     & \textbf{Corr} & \textbf{p-value} \\
\hline
Disorganized         & 0.318      & 0.0312         \\
Sympathetic          & -0.317     & 0.0321         \\
\hline
\end{tabular}
\caption{Significant viewer traits and ratings (p-value < 0.05) on Actions Scores}
\label{tab:sig-test-viewer-actions}
\vspace{-0.5mm}
\end{table}

\section{Analysis of Video Characteristics}

To answer question \textbf{RQ2}, the following information is collected from raters:

\textbf{Overall Impression:} raters' overall impression of the videos, including whether they find them boring, lively, persuasive, trustworthy, logical, professionally created, and making a valid point (see Question 2, 8 and 15 in Table \ref{tab:survey-video-a} and \ref{tab:survey-video-b}). Each response is converted into a score from [-1, 1].

\textbf{Arousing Emotions:} the emotions raters feel when watching the videos, including Ekman's 6 emotions \cite{ekman1971constants} and confused (see Question 12 in Table \ref{tab:survey-video-b}). Each emotion is scored 1 if selected present, and 0 otherwise.

\textbf{Speaker Characteristics:} the traits of the speakers appearing in videos. We select a subset of speaker traits used in \cite{yang2020makes} to define the level of charisma of a speaker, including charismatic, confident, eloquent, enthusiastic, intelligent, convincing, tough, charming, and angry (see Question 10 in Table \ref{tab:survey-video-b}). Each rating is converted into a score ranging [-1, 1].

We perform significance tests on the Pearson's correlation between the above traits and ratings and the three metric scores. For our Enjoyment Score, the significant results are listed in Table \ref{tab:sig-test-video-enjoy}. For pro-QAnon videos, those rated as more valid and persuasive are enjoyed more by viewers. However, no other significant traits are found to be associated with the Enjoyment Score of anti-QAnon videos, or of all QAnon videos in general. 

\begin{table}[!htb]
\small
\centering
\begin{tabular}{ccc}
\hline
\multicolumn{3}{c}{\textbf{Enjoyment on Pro-QAnon Videos}} \\
\textbf{Feature}  & \textbf{Corr} & \textbf{p-value} \\
\hline
Validness         & 0.999      & 0.0234         \\
Persuasive        & 0.997      & 0.0452         \\
\hline
\end{tabular}
\caption{Significant video traits and ratings (p-value < 0.05) on the Enjoyment Scores}
\label{tab:sig-test-video-enjoy}
\vspace{-0.5mm}
\end{table}

Since the Content Score is a sum of persuasive, trustworthy, logical, and professional scores, we exclude these 4 traits when performing another set of correlation significance tests on our Content Score. As shown in Table \ref{tab:sig-test-video-content}, for anti-QAnon videos, if viewers feel disgusted or boredom when watching them, they tend to disagree with the content. No other significant traits are found to be specifically associated with the Content Score of pro-QAnon videos, or all selected QAnon videos in general.

\begin{table}[!htb]
\small
\centering
\begin{tabular}{ccc}
\hline
\multicolumn{3}{c}{\textbf{Content of Anti-QAnon Videos}} \\
\textbf{Feature} & \textbf{Corr} & \textbf{p-value} \\
\hline
Disgust          & -0.998     & 0.0440         \\
Boring           & -0.998     & 0.0440         \\
\hline
\end{tabular}
\caption{Significant video traits and ratings (p-value < 0.05) on the Content Scores}
\label{tab:sig-test-video-content}
\vspace{-0.5mm}
\end{table}

Looking at our Actions Scores, we find that viewer ratings that are positively correlated with supporting actions are whether the videos are trustworthy, persuasive, logical, and making a valid point. Similarly, for anti-QAnon videos, viewers are also more likely to take supporting actions after watching the videos if they think the videos are trustworthy. On the other hand, if the speakers in the videos are rated as enthusiastic, the viewers indicate that they are less likely to take supporting actions. For anti-QAnon videos, the liveliness of videos is also negatively correlated with supporting activity. No significant traits are found to be associated with the Actions Scores of pro-QAnon videos.

\begin{table}[!htb]
\small
\centering
\begin{tabular}{ccc}
\hline
\multicolumn{3}{c}{\textbf{Actions Likely after All Videos}}  \\
\textbf{Feature}   & \textbf{Corr}  & \textbf{p-value} \\
\hline
Trustworthy        & 0.968       & 0.00150         \\
Validness          & 0.964       & 0.00191         \\
Persuasive         & 0.905       & 0.0131          \\
Logical            & 0.875       & 0.0225          \\
Enthusiastic       & -0.951      & 0.0486          \\
\hline\hline
\multicolumn{3}{c}{\textbf{Actions after Anti Videos}} \\
\textbf{Feature}   & \textbf{Corr}  & \textbf{p-value} \\
\hline
Trustworthy        & 1.00       & 0.0114         \\
Lively             & -1.00      & 0.0167         \\
\hline
\end{tabular}
\caption{Significant video ratings (p-value < 0.05) on the Actions Scores}
\label{tab:sig-test-video-actions}
\vspace{-0.5mm}
\end{table}

\section{Multimodal Feature Analysis}
To answer \textbf{RQ3}, we further analyze multimodal features of these videos, including textual, acoustic, and visual features. We perform analysis on 2 levels: \textbf{(1)} inter-pausal unit (IPU) segment level; \textbf{(2)} whole video level. We further perform significance tests on the Pearson's correlation between all the multimodal features and the three metric scores on both IPU segment level and video level. The complete lists of significant multimodal features are summarized in Appendix \ref{appendix:sig-tests}, and here we highlight some of the key and interesting findings.

\subsection{Textual Features}
\label{subsec:text}
To obtain textual features, we extract speech transcripts of these videos using the Google Speech-to-Text service \footnote{\href{https://cloud.google.com/speech-to-text}{https://cloud.google.com/speech-to-text}}. We then use Linguistic Inquiry and Word Count (LIWC) \cite{pennebaker2015development} to extract lexico-semantic features, Grievance Dictionary \cite{van2021grievance} to extract psycholinguistic features, and VADER \cite{hutto2014vader} to extract textual sentiment scores. 

The list of significant segment level textual features are summarized in Table \ref{tab:sig-test-seg-text-enjoy}, \ref{tab:sig-test-seg-text-content}, and \ref{tab:sig-test-seg-text-actions} in Appendix \ref{appendix:sig-tests-seg-text}. In general, lexicons related to friends and gender are positively correlated with how viewers perceive the videos, in terms of how they enjoy watching the videos, agree with the content, and take active actions afterwards. Lexicons related to violence are negatively correlated with how viewers enjoy watching the videos. For pro-QAnon videos, lexicons related to violence, weapons, threat, power, and soldiers are significantly and negatively correlated with how viewers perceive the videos. These are when the topics such as war and crimes are being talked about. In addition, VADER sentiment is positively correlated with with how viewers perceive the videos for pro-QAnon videos. For anti-QAnon videos, lexicons related to friends are positively correlated with viewer perception.

On video level, no significant lexical features stands out for pro-QAnon videos. For anti-QAnon videos, or QAnon videos in general, lexicons related to loneliness positively affect how viewers enjoy watching the videos. Viewers also tend to agree with the content more if lexicons related to gender and family are mentioned; and they tend to disagree with the content if paranoia words such as "crazy" are mentioned. The complete list of significant video level textual features are summarized in Table \ref{tab:sig-test-video-text-enjoy}, \ref{tab:sig-test-video-text-content} and \ref{tab:sig-test-video-text-actions} in Appendix \ref{appendix:sig-tests-video-text}.

\subsection{Acoustic Features}
\label{subsec:acoustic}
We extract acoustic-prosodic features, such as pitch and intensity, because they are proven to be relevant to how people express emotion \cite{sudhakar2015analysis}, and attempt to be persuasive \cite{nguyen2021acoustic} and charismatic \cite{yang2020makes}. We also extract emotions from the videos' speech using SpeechBrain system \cite{speechbrain}.

The significant segment level acoustic features are listed in Table \ref{tab:sig-test-seg-audio-enjoy}, \ref{tab:sig-test-seg-audio-content}, and \ref{tab:sig-test-seg-audio-actions} in Appendix \ref{appendix:sig-tests-seg-audio}. In general, intensity and maximum pitch are negatively correlated with viewer's perception -- the louder the speakers are, the less likely that the viewers would enjoy the videos and the content. This is what we observe for all videos, including pro- and anti-QAnon videos. In addition, the more angry the speakers are, the less likely that the viewers would agree with the content.

\subsection{Visual Features}
\label{subsec:visual}
For visual features, we extract frame-level facial expression features with a pre-trained FER model \footnote{\href{https://github.com/WuJie1010/Facial-Expression-Recognition.Pytorch.git}{Facial-Expression-Recognition.Pytorch}}. We also detect weapons that appear in the videos using Clarifai's weapon detector model \footnote{\href{https://www.clarifai.com/models/weapon-detector}{Clarifai Weapon Detector}}, as we have proven in Secion \ref{subsec:text} that topics related to violence and war are correlated with viewer' perception.

The significant segment level visual features are listed in Table \ref{tab:sig-test-seg-visual-enjoy}, \ref{tab:sig-test-seg-visual-content}, and \ref{tab:sig-test-seg-visual-actions} in Appendix \ref{appendix:sig-tests-seg-visual}. In general, if speakers appear in the videos show surprise or sad facial expressions, viewers tend to have negative perception. However, speakers' angry expressions are positively correlated with viewers' perception. For anti-QAnon videos, speakers' negative expressions, such as fear and disgust,  are negatively correlated with how viewers would enjoy and agree with the videos. In addition, the appearance of weapons, regardless of what types of weapons, has a negative impact on viewers' perception. This agrees with what we observe in textual features, where words related to violence are negatively correlated with viewer' perception.

Similarly, on video level, we observe that speakers' surprise and fear expressions are negatively correlated with how viewers perceive the videos. The complete list of significant video level visual features are summarized in Table \ref{tab:sig-test-video-visual-enjoy}, \ref{tab:sig-test-video-visual-content}, and \ref{tab:sig-test-video-visual-actions}.

\section{Conclusions and Future Work}
In this study, we have collected a corpus of QAnon videos and have designed a comprehensive questionnaire. With the responses we collect from the questionnaire, we are able to propose 3 metrics to evaluate viewers' perception towards the videos, and outline the traits of viewers that QAnon videos are the most appealing to, including their personalities, media consumption, and presumption about other radicalizing groups. In addition, we identify video characteristics, including generic content traits and arousing emotions, that impact viewers' perception of the videos.

In future, we will analyze multimodal features to investigate what modality features contribute to viewers' perception. We also aim to utilize multimodal features to build models for identifying radical content and techniques.

\section*{Limitations}
One of the limitations of this study is the unbalanced distribution of rater demographics. Specifically, 91\% of our raters report themselves having a Bachelor's degree or higher, 84.7\% of the raters consider themselves to be liberal and moderate, and 91\% of the raters belong to the 18-29 age group. In future, we will collect crowdsourcing annotations from a more diverse population.

Another limitation of our study is the size of the data we put out with the questionnaire -- 6 videos with 3 pro- and 3 anti-QAnon, because manually selecting videos that are the most relevant and appropriate is extremely time-consuming. However, with this study as an initial step, we will utilize the conclusions we have drawn and aim to make use of the full corpus of 5,924 QAnon videos that we have collected so far for further analysis and model building.

\section*{Ethics Statement}
We discuss the ethical considerations of our study as follows:

\textbf{Data Collection:} We collect videos from YouTube and BitChute, where all videos and their associated metadata are available to public. For YouTube videos, we use the official Google Developer API\footnote{\href{https://developers.google.com/youtube/v3/docs}{https://developers.google.com/youtube/v3/docs}}. For BitChute videos, we scrape publicly available videos and data without utilizing any internal APIs and private access.

\textbf{Questionnaire Response Collection:} All raters take part in the questionnaire participate voluntarily and are fully aware of any risks of harm associated with their participation. We do not collect any personal information that would allow us to identify the raters, or to associate them with their responses.

\textbf{Data Release:} Due to the sensitivity of the data, the raw videos, video metadata, and detailed questionnaire responses are not made available yet on any platforms. However, we are willing to consider sharing them with other research groups upon request.

\newpage
\bibliography{anthology,references}
\bibliographystyle{acl_natbib}

\newpage
\appendix

\section{Significant Multimodal Features}
\label{appendix:sig-tests}

\subsection{Segment Level Significant Features}
\label{appendix:sig-tests-seg}

\subsubsection{Textual Features}
\label{appendix:sig-tests-seg-text}

\begin{table}[!htb]
\small
\centering
\begin{tabular}{ccc}
\hline
\multicolumn{3}{c}{\textbf{Enjoyment on All Videos}}        \\
\textbf{Feature}     & \textbf{Corr}   & \textbf{p-value}   \\
\hline
violence             & -0.138          & 0.0247             \\
deadline             & -0.125          & 0.0429             \\
i                    & 0.165           & 0.00716            \\
they                 & -0.141          & 0.0220             \\
male                 & 0.122           & 0.0475             \\
social               & -0.122          & 0.0478             \\
negate               & 0.122           & 0.0488             \\
\hline\hline
\multicolumn{3}{c}{\textbf{Enjoyment on Pro-QAnon Videos}}  \\
\textbf{Feature}     & \textbf{Corr}   & \textbf{p-value}   \\
\hline
sentiment            & 0.205           & 0.0337             \\
weaponry             & -0.387          & 0.0000384          \\
violence             & -0.324          & 0.000671           \\
god                  & -0.266          & 0.00556            \\
soldier              & -0.211          & 0.0294             \\
threat               & -0.202          & 0.0370             \\
focuspresent         & 0.377           & 0.0000630          \\
they                 & -0.351          & 0.000215           \\
power                & -0.328          & 0.000554           \\
ipron                & 0.323           & 0.000677           \\
cogproc              & 0.301           & 0.00162            \\
auxverb              & 0.291           & 0.00235            \\
negate               & 0.280           & 0.00351            \\
we                   & -0.277          & 0.00388            \\
social               & -0.273          & 0.00451            \\
affiliation          & -0.267          & 0.00543            \\
i                    & 0.261           & 0.00664            \\
tentat               & 0.260           & 0.00675            \\
negemo               & -0.242          & 0.0120             \\
drives               & -0.241          & 0.0122             \\
adverb               & 0.231           & 0.0167             \\
ppron                & -0.228          & 0.0182             \\
anger                & -0.222          & 0.0215             \\
verb                 & 0.219           & 0.0233             \\
informal             & 0.219           & 0.0235             \\
differ               & 0.212           & 0.0281             \\
health               & -0.210          & 0.0300             \\
body                 & -0.209          & 0.0310             \\
discrep              & -0.208          & 0.0314             \\
bio                  & -0.196          & 0.0428             \\
quant                & 0.191           & 0.0484             \\
\hline\hline
\multicolumn{3}{c}{\textbf{Enjoyment on Anti-QAnon Videos}} \\
\textbf{Feature}     & \textbf{Corr}   & \textbf{p-value}   \\
\hline
interrog             & -0.165          & 0.0395             \\
\hline
\end{tabular}
\caption{Significant segment level textual features (p-value \textless 0.05) on Enjoyment Scores}
\label{tab:sig-test-seg-text-enjoy}
\vspace{-0.5mm}
\end{table}

\begin{table}[!htb]
\small
\centering
\begin{tabular}{ccc}
\hline
\multicolumn{3}{c}{\textbf{Content of All Videos}}        \\
\textbf{Feature}    & \textbf{Corr}   & \textbf{p-value}  \\
\hline
god                 & -0.128          & 0.0376            \\
time                & -0.217          & 0.000393          \\
differ              & 0.139           & 0.0245            \\
friend              & 0.135           & 0.0285            \\
insight             & -0.134          & 0.0303            \\
ingest              & 0.122           & 0.0475            \\
\hline\hline
\multicolumn{3}{c}{\textbf{Content of Pro-QAnon Videos}}  \\
\textbf{Feature}    & \textbf{Corr}   & \textbf{p-value}  \\
\hline
sentiment           & 0.205           & 0.0337            \\
weaponry            & -0.387          & 0.0000384         \\
violence            & -0.324          & 0.000671          \\
god                 & -0.266          & 0.00556           \\
soldier             & -0.211          & 0.0294            \\
threat              & -0.202          & 0.0370            \\
focuspresent        & 0.377           & 0.0000630         \\
they                & -0.351          & 0.000215          \\
power               & -0.328          & 0.000554          \\
ipron               & 0.323           & 0.000677          \\
cogproc             & 0.301           & 0.00162           \\
auxverb             & 0.291           & 0.00235           \\
negate              & 0.280           & 0.00351           \\
we                  & -0.277          & 0.00388           \\
social              & -0.273          & 0.00451           \\
affiliation         & -0.267          & 0.00543           \\
i                   & 0.261           & 0.00664           \\
tentat              & 0.260           & 0.00675           \\
negemo              & -0.242          & 0.0120            \\
drives              & -0.241          & 0.0122            \\
adverb              & 0.231           & 0.0167            \\
ppron               & -0.228          & 0.0182            \\
anger               & -0.222          & 0.0215            \\
verb                & 0.219           & 0.0233            \\
informal            & 0.219           & 0.0235            \\
differ              & 0.212           & 0.0281            \\
health              & -0.210          & 0.0300            \\
body                & -0.209          & 0.0310            \\
discrep             & -0.208          & 0.0314            \\
bio                 & -0.196          & 0.0428            \\
quant               & 0.191           & 0.0484            \\
\hline\hline
\multicolumn{3}{c}{\textbf{Content of Anti-QAnon Videos}} \\
\textbf{Feature}    & \textbf{Corr}   & \textbf{p-value}  \\
\hline
time                & -0.285          & 0.000316          \\
friend              & 0.213           & 0.00765           \\
focuspast           & -0.164          & 0.0409            \\
female              & 0.163           & 0.0420            \\
ingest              & 0.163           & 0.0424            \\
conj                & -0.157          & 0.0498            \\
\hline
\end{tabular}
\caption{Significant segment level textual features (p-value \textless 0.05) on Content Scores}
\label{tab:sig-test-seg-text-content}
\end{table}

\begin{table}[!htb]
\small
\centering
\begin{tabular}{ccc}
\hline
\multicolumn{3}{c}{\textbf{Actions after All Videos}}        \\
\textbf{Feature}     & \textbf{Corr}    & \textbf{p-value}   \\
\hline
time                 & -0.214           & 0.000479           \\
friend               & 0.147            & 0.0173             \\
insight              & -0.132           & 0.0321             \\
negate               & 0.130            & 0.0353             \\
female               & 0.130            & 0.0353             \\
ingest               & 0.126            & 0.0417             \\
\hline\hline
\multicolumn{3}{c}{\textbf{Actions after Pro-QAnon Videos}}  \\
\textbf{Feature}     & \textbf{Corr}    & \textbf{p-value}   \\
\hline
sentiment            & 0.205            & 0.0337             \\
weaponry             & -0.387           & 0.0000384          \\
violence             & -0.324           & 0.000671           \\
god                  & -0.266           & 0.00556            \\
soldier              & -0.211           & 0.0294             \\
threat               & -0.202           & 0.0370             \\
focuspresent         & 0.377            & 0.0000630          \\
they                 & -0.351           & 0.000215           \\
power                & -0.328           & 0.000554           \\
ipron                & 0.323            & 0.000677           \\
cogproc              & 0.301            & 0.00162            \\
auxverb              & 0.291            & 0.00235            \\
negate               & 0.280            & 0.00351            \\
we                   & -0.277           & 0.00388            \\
social               & -0.273           & 0.00451            \\
affiliation          & -0.267           & 0.00543            \\
i                    & 0.261            & 0.00664            \\
tentat               & 0.260            & 0.00675            \\
negemo               & -0.242           & 0.0120             \\
drives               & -0.241           & 0.0122             \\
adverb               & 0.231            & 0.0167             \\
ppron                & -0.228           & 0.0182             \\
anger                & -0.222           & 0.0215             \\
verb                 & 0.219            & 0.0233             \\
informal             & 0.219            & 0.0235             \\
differ               & 0.212            & 0.0281             \\
health               & -0.210           & 0.0300             \\
body                 & -0.209           & 0.0310             \\
discrep              & -0.208           & 0.0314             \\
bio                  & -0.196           & 0.0428             \\
quant                & 0.191            & 0.0484             \\
\hline\hline
\multicolumn{3}{c}{\textbf{Actions after Anti-QAnon Videos}} \\
\textbf{Feature}     & \textbf{Corr}    & \textbf{p-value}   \\
\hline
time                 & -0.266           & 0.000795           \\
friend               & 0.195            & 0.0146             \\
insight              & -0.161           & 0.0442             \\ 
\hline
\end{tabular}
\caption{Significant segment level textual features (p-value \textless{} 0.05) on Actions Scores}
\label{tab:sig-test-seg-text-actions}
\end{table}

\clearpage
\subsubsection{Acoustic Features}
\label{appendix:sig-tests-seg-audio}

\begin{table}[!htb]
\small
\centering
\begin{tabular}{ccc}
\hline
\multicolumn{3}{c}{\textbf{Enjoyment on All Videos}}        \\
\textbf{Feature}     & \textbf{Corr}   & \textbf{p-value}   \\
\hline
Max Intensity        & -0.660          & 3.14E-34           \\
Mean Intensity       & -0.654          & 1.55E-33           \\
Sd Intensity         & -0.565          & 1.32E-23           \\
Sd Pitch             & -0.361          & 1.68E-09           \\
Max Pitch            & -0.354          & 3.68E-09           \\
Jitter               & 0.303           & 5.66E-07           \\
Mean Pitch           & 0.230           & 0.000164           \\
Shimmer              & -0.134          & 0.0301             \\
\hline\hline
\multicolumn{3}{c}{\textbf{Enjoyment on Pro-QAnon Videos}}  \\
\textbf{Feature}     & \textbf{Corr}   & \textbf{p-value}   \\
\hline
HNR                  & 0.870           & 5.76E-34           \\
Mean Pitch           & 0.738           & 1.26E-19           \\
Mean Intensity       & -0.713          & 7.18E-18           \\
Jitter               & 0.649           & 4.15E-14           \\
Shimmer              & -0.640          & 1.17E-13           \\
Min Pitch            & 0.562           & 2.97E-10           \\
Max Intensity        & -0.507          & 2.46E-08           \\
Sd Pitch             & -0.440          & 2.12E-06           \\
Max Pitch            & -0.424          & 5.42E-06           \\
Min Intensity        & -0.329          & 0.000548           \\
Sd Intensity         & -0.230          & 0.0169             \\
\hline\hline
\multicolumn{3}{c}{\textbf{Enjoyment on Anti-QAnon Videos}} \\
\textbf{Feature}     & \textbf{Corr}   & \textbf{p-value}   \\
\hline
Max Intensity        & -0.832          & 3.02E-41           \\
Mean Intensity       & -0.829          & 9.96E-41           \\
Sd Intensity         & -0.678          & 2.25E-22           \\
Sd Pitch             & -0.348          & 8.33E-06           \\
Max Pitch            & -0.336          & 0.0000184          \\
HNR                  & -0.328          & 0.0000285          \\
Min Intensity        & 0.298           & 0.000161           \\
Jitter               & 0.172           & 0.0316             \\
\hline
\end{tabular}
\caption{Significant segment level acoustic features (p-value \textless 0.05) on Enjoyment Scores}
\label{tab:sig-test-seg-audio-enjoy}
\end{table}

\begin{table}[!htb]
\small
\centering
\begin{tabular}{ccc}
\hline
\multicolumn{3}{c}{\textbf{Content of All Videos}}        \\
\textbf{Feature}    & \textbf{Corr}   & \textbf{p-value}  \\
\hline
anger               & -0.169          & 0.00602           \\
Min Intensity       & 0.618           & 4.36E-29          \\
Sd Intensity        & -0.428          & 3.71E-13          \\
Mean Intensity      & 0.367           & 8.08E-10          \\
Max Intensity       & 0.353           & 4.06E-09          \\
HNR                 & -0.234          & 0.000129          \\
Min Pitch           & 0.192           & 0.00171           \\
\hline\hline
\multicolumn{3}{c}{\textbf{Content of Pro-QAnon Videos}}  \\
\textbf{Feature}    & \textbf{Corr}   & \textbf{p-value}  \\
\hline
HNR                 & 0.870           & 5.76E-34          \\
Mean Pitch          & 0.738           & 1.26E-19          \\
Mean Intensity      & -0.713          & 7.18E-18          \\
Jitter              & 0.649           & 4.15E-14          \\
Shimmer             & -0.640          & 1.17E-13          \\
Min Pitch           & 0.562           & 2.97E-10          \\
Max Intensity       & -0.507          & 2.46E-08          \\
Sd Pitch            & -0.440          & 2.12E-06          \\
Max Pitch           & -0.424          & 5.42E-06          \\
Min Intensity       & -0.329          & 0.000548          \\
Sd Intensity        & -0.230          & 0.0169            \\
\hline\hline
\multicolumn{3}{c}{\textbf{Content of Anti-QAnon Videos}} \\
\textbf{Feature}    & \textbf{Corr}   & \textbf{p-value}  \\
\hline
Min Intensity       & 0.676           & 3.58E-22          \\
Sd Intensity        & -0.419          & 5.21E-08          \\
HNR                 & -0.311          & 0.0000767         \\
Mean Intensity      & 0.179           & 0.0256            \\
\hline
\end{tabular}
\caption{Significant segment level acoustic features (p-value \textless 0.05) on Content Scores}
\label{tab:sig-test-seg-audio-content}
\end{table}

\begin{table}[!htb]
\small
\centering
\begin{tabular}{ccc}
\hline
\multicolumn{3}{c}{\textbf{Actions after All Videos}}        \\
\textbf{Feature}     & \textbf{Corr}    & \textbf{p-value}   \\
\hline
Sd Intensity         & -0.543           & 1.33E-21           \\
Min Intensity        & 0.518            & 1.99E-19           \\
Max Pitch            & -0.173           & 0.00488            \\
Sd Pitch             & -0.173           & 0.00494            \\
HNR                  & -0.164           & 0.00767            \\
\hline\hline
\multicolumn{3}{c}{\textbf{Actions after Pro-QAnon Videos}}  \\
\textbf{Feature}     & \textbf{Corr}    & \textbf{p-value}   \\
\hline
HNR                  & 0.870            & 5.76E-34           \\
Mean Pitch           & 0.738            & 1.26E-19           \\
Mean Intensity       & -0.713           & 7.18E-18           \\
Jitter               & 0.649            & 4.15E-14           \\
Shimmer              & -0.640           & 1.17E-13           \\
Min Pitch            & 0.562            & 2.97E-10           \\
Max Intensity        & -0.507           & 2.46E-08           \\
Sd Pitch             & -0.440           & 2.12E-06           \\
Max Pitch            & -0.424           & 5.42E-06           \\
Min Intensity        & -0.329           & 0.000548           \\
Sd Intensity         & -0.230           & 0.0169             \\
\hline\hline
\multicolumn{3}{c}{\textbf{Actions after Anti-QAnon Videos}} \\
\textbf{Feature}     & \textbf{Corr}    & \textbf{p-value}   \\
\hline
Min Intensity        & 0.687            & 3.77E-23           \\
Sd Intensity         & -0.569           & 8.67E-15           \\
HNR                  & -0.372           & 1.74E-06           \\
Max Pitch            & -0.164           & 0.0409             \\
\hline
\end{tabular}
\caption{Significant segment level acoustic features (p-value \textless 0.05) on Actions Scores}
\label{tab:sig-test-seg-audio-actions}
\end{table}

\clearpage
\subsubsection{Visual Features}
\label{appendix:sig-tests-seg-visual}

\begin{table}[!htb]
\small
\centering
\begin{tabular}{ccc}
\hline
\multicolumn{3}{c}{\textbf{Enjoyment on All Videos}}        \\
\textbf{Feature}     & \textbf{Corr}   & \textbf{p-value}   \\
\hline
neutral              & -0.270          & 1.23E-10           \\
surprise             & -0.143          & 7.95E-04           \\
happy                & 0.126           & 3.20E-03           \\
sad                  & -0.117          & 6.23E-03           \\
has\_weapon          & -0.215          & 1.01E-06           \\
long-gun             & -0.210          & 1.74E-06           \\
sword                & -0.148          & 0.000799           \\
\hline\hline
\multicolumn{3}{c}{\textbf{Enjoyment on Pro-QAnon Videos}}  \\
\textbf{Feature}     & \textbf{Corr}   & \textbf{p-value}   \\
\hline
happy                & 0.259           & 0.0000105          \\
neutral              & -0.234          & 0.0000722          \\
sad                  & -0.226          & 0.000127           \\
angry                & 0.166           & 0.00532            \\
surprise             & -0.143          & 0.0160             \\
has\_weapon          & -0.243          & 0.000133           \\
long-gun             & -0.220          & 0.000567           \\
sword                & -0.184          & 0.00413            \\
\hline\hline
\multicolumn{3}{c}{\textbf{Enjoyment on Anti-QAnon Videos}} \\
\textbf{Feature}     & \textbf{Corr}   & \textbf{p-value}   \\
\hline
fear                 & -0.230          & 0.000154           \\
surprise             & -0.169          & 0.00579            \\
disgust              & -0.156          & 0.0108             \\
\hline
\end{tabular}
\caption{Significant segment level visual features (p-value \textless 0.05) on Enjoyment Scores}
\label{tab:sig-test-seg-visual-enjoy}
\end{table}

\begin{table}[!htb]
\small
\centering
\begin{tabular}{ccc}
\hline
\multicolumn{3}{c}{\textbf{Content of All Videos}}        \\
\textbf{Feature}    & \textbf{Corr}   & \textbf{p-value}  \\
\hline
angry               & 0.311           & 9.41E-14          \\
sad                 & -0.169          & 0.0000726         \\
surprise            & -0.117          & 0.00628           \\
happy               & 0.107           & 0.0122            \\
neutral             & -0.0995         & 0.0198            \\
long-gun            & -0.139          & 0.00163           \\
has\_weapon         & -0.0923         & 0.0376            \\
sword               & -0.0904         & 0.0418            \\
\hline\hline
\multicolumn{3}{c}{\textbf{Content of Pro-QAnon Videos}}  \\
\textbf{Feature}    & \textbf{Corr}   & \textbf{p-value}  \\
\hline
happy               & 0.259           & 0.0000109         \\
neutral             & -0.235          & 0.0000673         \\
sad                 & -0.226          & 0.000129          \\
angry               & 0.165           & 0.00548           \\
surprise            & -0.143          & 0.0160            \\
has\_weapon         & -0.243          & 0.000136          \\
long-gun            & -0.220          & 0.000561          \\
sword               & -0.184          & 0.00409           \\
\hline\hline
\multicolumn{3}{c}{\textbf{Content of Anti-QAnon Videos}} \\
\textbf{Feature}    & \textbf{Corr}   & \textbf{p-value}  \\
\hline
angry               & 0.482           & 7.06E-17          \\
neutral             & 0.167           & 0.00647           \\
fear                & -0.123          & 0.0451            \\
has\_weapon         & 0.141           & 0.0213            \\
long-gun            & 0.143           & 0.0197            \\
\hline
\end{tabular}
\caption{Significant segment level visual features (p-value \textless 0.05) on Content Scores}
\label{tab:sig-test-seg-visual-content}
\end{table}

\begin{table}[!htb]
\small
\centering
\begin{tabular}{ccc}
\hline
\multicolumn{3}{c}{\textbf{Actions after All Videos}}        \\
\textbf{Feature}     & \textbf{Corr}    & \textbf{p-value}   \\
\hline
angry                & 0.312            & 8.15E-14           \\
sad                  & -0.148           & 0.000514           \\
surprise             & -0.131           & 0.00215            \\
neutral              & -0.0905          & 0.0342             \\
long-gun             & -0.124           & 0.00526            \\
\hline\hline
\multicolumn{3}{c}{\textbf{Actions after Pro-QAnon Videos}}  \\
\textbf{Feature}     & \textbf{Corr}    & \textbf{p-value}   \\
\hline
happy                & 0.268            & 5.20E-06           \\
sad                  & -0.231           & 0.0000892          \\
neutral              & -0.212           & 0.000343           \\
angry                & 0.178            & 0.00270            \\
surprise             & -0.141           & 0.0182             \\
has\_weapon          & -0.251           & 0.0000816          \\
long-gun             & -0.215           & 0.000746           \\
sword                & -0.179           & 0.00531            \\
\hline\hline
\multicolumn{3}{c}{\textbf{Actions after Anti-QAnon Videos}} \\
\textbf{Feature}     & \textbf{Corr}    & \textbf{p-value}   \\
\hline
angry                & 0.429            & 2.43E-13           \\
fear                 & -0.179           & 0.00335            \\
neutral              & 0.128            & 0.0374             \\
surprise             & -0.121           & 0.0495             \\
has\_weapon          & 0.125            & 0.0424             \\
\hline
\end{tabular}
\caption{Significant segment level visual features (p-value \textless 0.05) on Actions Scores}
\label{tab:sig-test-seg-visual-actions}
\end{table}

\clearpage
\subsection{Video Level Significant Features}
\label{appendix:sig-tests-video}

\subsubsection{Textual Features}
\label{appendix:sig-tests-video-text}

\begin{table}[!htb]
\small
\centering
\begin{tabular}{ccc}
\hline
\multicolumn{3}{c}{\textbf{Enjoyment on All Videos}}        \\
\textbf{Feature}     & \textbf{Corr}   & \textbf{p-value}   \\
\hline
loneliness           & 0.969           & 0.00645            \\
planning             & -0.921          & 0.0265             \\
\hline\hline
\multicolumn{3}{c}{\textbf{Enjoyment on Anti-QAnon Videos}} \\
\textbf{Feature}     & \textbf{Corr}   & \textbf{p-value}   \\
\hline
loneliness           & 0.998           & 0.0360             \\
honour               & 0.998           & 0.0429             \\
home                 & 1.000           & 0.00794            \\
\hline
\end{tabular}
\caption{Significant video level textual features (p-value \textless 0.05) on Enjoyment Scores}
\label{tab:sig-test-video-text-enjoy}
\end{table}

\begin{table}[!htb]
\small
\centering
\begin{tabular}{ccc}
\hline
\multicolumn{3}{c}{\textbf{Content of All Videos}}        \\
\textbf{Feature}   & \textbf{Corr}   & \textbf{p-value}   \\
\hline
relativ            & -0.933          & 0.0208             \\
time               & -0.903          & 0.0358             \\
percept            & -0.899          & 0.0380             \\
sexual             & 0.889           & 0.0436             \\
adj                & 0.885           & 0.0462             \\
\hline
\multicolumn{3}{c}{\textbf{Content of Anti-QAnon Videos}} \\
\textbf{Feature}   & \textbf{Corr}   & \textbf{p-value}   \\
\hline\hline
god                & -1.000          & 0.0174             \\
paranoia           & -1.000          & 0.0174             \\
family             & 0.999           & 0.0234             \\
female             & 0.998           & 0.0440             \\
sexual             & 0.998           & 0.0440             \\
ingest             & 0.998           & 0.0440             \\
death              & 0.998           & 0.0440             \\
swear              & 0.998           & 0.0440             \\
\hline
\end{tabular}
\caption{Significant video level textual features (p-value \textless 0.05) on Content Scores}
\label{tab:sig-test-video-text-content}
\end{table}

\begin{table}[!htb]
\small
\centering
\begin{tabular}{ccc}
\hline
\multicolumn{3}{c}{\textbf{Actions after All Videos}}        \\
\textbf{Feature}     & \textbf{Corr}    & \textbf{p-value}   \\
\hline
relativ              & -0.992           & 0.000852           \\
adj                  & 0.950            & 0.0135             \\
time                 & -0.942           & 0.0165             \\
percept              & -0.924           & 0.0250             \\
hear                 & -0.907           & 0.0337             \\
ingest               & 0.885            & 0.0458             \\
\hline\hline
\multicolumn{3}{c}{\textbf{Actions after Anti-QAnon Videos}} \\
\textbf{Feature}     & \textbf{Corr}    & \textbf{p-value}   \\
\hline
help                 & 1.000            & 0.0146             \\
percept              & -1.000           & 0.00369            \\
compare              & 1.000            & 0.0103             \\
you                  & -0.999           & 0.0271             \\
relativ              & -0.998           & 0.0364             \\
bio                  & 0.997            & 0.0488             \\
\hline
\end{tabular}
\caption{Significant video level textual features (p-value \textless 0.05) on Actions Scores}
\label{tab:sig-test-video-text-actions}
\end{table}

\newpage
\subsubsection{Acoustic Features}
\label{appendix:sig-tests-video-audio}

\begin{table}[!htb]
\small
\centering
\begin{tabular}{ccc}
\hline
\multicolumn{3}{c}{\textbf{Enjoyment on All Videos}}        \\
\textbf{Feature}     & \textbf{Corr}   & \textbf{p-value}   \\
\hline
neutral              & 0.909           & 0.0323             \\
Sd Pitch             & -0.916          & 0.0288             \\
Max Pitch            & -0.916          & 0.0291             \\
Sd Intensity         & -0.884          & 0.0467             \\
\hline\hline
\multicolumn{3}{c}{\textbf{Enjoyment on Anti-QAnon Videos}} \\
\textbf{Feature}     & \textbf{Corr}   & \textbf{p-value}   \\
\hline
Max Pitch            & -1.00           & 0.0112             \\
Sd Pitch             & -1.00           & 0.0494             \\
\hline
\end{tabular}
\caption{Significant video level acoustic features (p-value \textless 0.05) on Enjoyment Scores}
\label{tab:sig-test-video-audio-enjoy}
\end{table}

\begin{table}[!htb]
\small
\centering
\begin{tabular}{ccc}
\hline
\multicolumn{3}{c}{\textbf{Actions after Anti-QAnon Videos}} \\
\textbf{Feature}    & \textbf{Corr}    & \textbf{p-value}    \\
\hline
Min Intensity       & 0.997            & 0.0500              \\
\hline
\end{tabular}
\caption{Significant video level acoustic features (p-value \textless 0.05) on Actions Scores}
\label{tab:sig-test-video-audio-actions}
\end{table}

\subsubsection{Visual Features}
\label{appendix:sig-tests-video-visual}

\begin{table}[!htb]
\small
\centering
\begin{tabular}{ccc}
\hline
\multicolumn{3}{c}{\textbf{Enjoyment on All Videos}} \\
\textbf{Feature}  & \textbf{Corr} & \textbf{p-value} \\
\hline
surprise          & -0.894        & 0.0163           \\
\hline
\end{tabular}
\caption{Significant video level visual features (p-value \textless 0.05) on Enjoyment Scores}
\label{tab:sig-test-video-visual-enjoy}
\end{table}

\begin{table}[!htb]
\small
\centering
\begin{tabular}{ccc}
\hline
\multicolumn{3}{c}{\textbf{Content of All Videos}}        \\
\textbf{Feature}   & \textbf{Corr}   & \textbf{p-value}   \\
\hline
surprise           & -0.821          & 0.0450             \\
\hline\hline
\multicolumn{3}{c}{\textbf{Content of Anti-QAnon Videos}} \\
\textbf{Feature}   & \textbf{Corr}   & \textbf{p-value}   \\
\hline
angry              & 1.000           & 0.00109            \\
fear               & -0.998          & 0.0361             \\
sword              & 0.998           & 0.0440             \\
\hline
\end{tabular}
\caption{Significant video level visual features (p-value \textless 0.05) on Content Scores}
\label{tab:sig-test-video-visual-content}
\end{table}

\begin{table}[!htb]
\small
\centering
\begin{tabular}{ccc}
\hline
\multicolumn{3}{c}{\textbf{Actions after All Videos}} \\
\textbf{Feature}  & \textbf{Corr}  & \textbf{p-value} \\
\hline
fear              & -0.812         & 0.0495           \\
\hline
\end{tabular}
\caption{Significant video level visual features (p-value \textless 0.05) on Actions Scores}
\label{tab:sig-test-video-visual-actions}
\end{table}

\clearpage
\onecolumn
\section{Questionnaire Questions}
\label{appendix:survey-questions}

\newcommand{\Qq}[1]{\textbf{#1}}
\newcommand{\Qo}{$\Box$}
\newcounter{qr}

\newcommand{\Qline}[1]{\noindent\rule{#1}{0.6pt}}

\newenvironment{Qlist}{%
\renewcommand{\labelitemi}{\Qo}
\begin{itemize}
\setlength\itemsep{-0.1em}
}{%
\end{itemize}
}

\newcommand{\Qitem}[1]{%
\begin{enumerate}[topsep=0pt]
\item[] #1
\end{enumerate}
}

{\renewcommand{\arraystretch}{0.8}
\begin{table}[htp]
\small
\begin{tabular}{p{15cm}}

\Qitem{ \Qq{1. What is your gender}
\begin{Qlist}
\item Male
\item Female
\item Nonbinary
\item Prefer not to say
\end{Qlist}
} \\[-2ex]

\Qitem{ \Qq{2. Which age group describes you?}
\begin{Qlist}
\item 18-29
\item 30-39
\item 40-49
\item 50-59
\item 60 or over
\end{Qlist}
} \\[-2ex]

\Qitem{ \Qq{3. What is your ethnicity?}
\begin{Qlist}
\item American Indian or Alaska Native
\item Asian
\item Black or African American
\item Native Hawaiian or Other Pacific Islander
\item White
\item Other
\end{Qlist}
} \\[-2ex]

\Qitem{ \Qq{4. What is the highest level of education you've completed?}
\begin{Qlist}
\item Some high school or less 
\item High school diploma
\item Associate's degree
\item Bachelor's degree
\item Master's degree
\item Doctorate degree
\end{Qlist}
} \\[-2ex]

\Qitem{ \Qq{5. Do you consider yourself to be conservative or liberal when thinking about politics?}
\begin{Qlist}
\item Conservative
\item Liberal
\item Moderate
\item Other/Undecided
\end{Qlist}
} \\[-2ex]

\Qitem{ \Qq{6. Here are a number of personality traits that may or may not apply to you. Please tick a number next to each statement to indicate the extent to which you agree or disagree with that statement. You should rate the extend to which the pair of traits applies to you, even if one characteristic applies more strongly than the other.}
\newline
\begin{tabular}{cccccc}
\textbf{} & \begin{tabular}[x]{@{}c@{}}1-Disagree\\strongly\end{tabular} & \begin{tabular}[x]{@{}c@{}}2-Disagree\\slightly\end{tabular} & \begin{tabular}[x]{@{}c@{}}3-Neither agree\\nor disagree\end{tabular} & \begin{tabular}[x]{@{}c@{}}4-Agree\\slightly\end{tabular} & \begin{tabular}[x]{@{}c@{}}5-Agree\\strongly\end{tabular} \\
Extraverted, enthusiastic & \Qo& \Qo& \Qo& \Qo& \Qo\\
Dependable, self-disciplined & \Qo& \Qo& \Qo& \Qo& \Qo\\
Anxious, easily upset & \Qo& \Qo& \Qo& \Qo& \Qo\\
Open to new experiences, complex & \Qo& \Qo& \Qo& \Qo& \Qo\\
Reserved, quiet & \Qo& \Qo& \Qo& \Qo& \Qo\\
Sympathetic, warm & \Qo& \Qo& \Qo& \Qo& \Qo\\
Disorganized, careless & \Qo& \Qo& \Qo& \Qo& \Qo\\
Calm, emotionally stable & \Qo& \Qo& \Qo& \Qo& \Qo\\
Conventional, uncreative & \Qo& \Qo& \Qo& \Qo& \Qo\\
\end{tabular}

}

\end{tabular}
\caption{Demographic Information}
\label{tab:survey-demographic}
\end{table}
}

{\renewcommand{\arraystretch}{0.8}
\begin{table*}[t]
\small
\begin{tabular}{p{15cm}}

\Qitem{ \Qq{1. What is your opinion of the following groups?}
\newline
\begin{tabular}{ccccc}
\textbf{} & Positive & Neutral & Negative & Never Heard of it  \\
QAnon & \Qo& \Qo& \Qo& \Qo \\
Antifa & \Qo& \Qo& \Qo& \Qo\\
Proud Boys & \Qo& \Qo& \Qo& \Qo\\
Oath Keepers & \Qo& \Qo& \Qo& \Qo\\
BLM & \Qo& \Qo& \Qo& \Qo \\
\end{tabular}
} \\

\Qitem{ \Qq{2. Do you trust the following media as credible sources?}
\newline
\begin{tabular}{ccccc}
\textbf{} & Positive & Neutral & Negative & Never Heard of it  \\
Fox News (foxnews.com) & \Qo & \Qo& \Qo& \Qo\\
Breitbart News (breitbart.com) & \Qo & \Qo& \Qo& \Qo\\
MSNBC News (msnbc.com) & \Qo & \Qo& \Qo& \Qo\\
PBS News (pbs.org) & \Qo & \Qo& \Qo& \Qo\\
Associated Press News (apnews.com) & \Qo & \Qo& \Qo& \Qo\\
NPR (npr.org) & \Qo & \Qo& \Qo& \Qo\\
The Wall Street Journal (wsj.com) & \Qo & \Qo& \Qo& \Qo\\
CNN (cnn.com) & \Qo & \Qo& \Qo& \Qo\\
\end{tabular}
} \\

\end{tabular}
\caption{Introductory Information}
\label{tab:survey-introductory}
\end{table*}
}

{\renewcommand{\arraystretch}{0.8}
\begin{table*}[t]
\small
\begin{tabular}{p{15cm}}

\Qitem{ \Qq{1. Did you understand the video?}
\begin{Qlist}
\item Yes
\item No
\end{Qlist}
} \\[-2ex]

\Qitem{ \Qq{2. Do you think the video was professionally produced with good quality?}
\begin{Qlist}
\item Yes
\item No
\end{Qlist}
} \\[-2ex]

\Qitem{ \Qq{3. Who do you think the video was trying to appeal to?}: \Qline{6cm} } \\[-2ex]

\Qitem{ \Qq{4. Was there any violence displayed in the video?}
\begin{Qlist}
\item Yes
\item No
\end{Qlist}
} \\[-2ex]

\Qitem{ \Qq{5. Was there any music in video?}
\begin{Qlist}
\item Yes
\item No
\end{Qlist}
} \\

\Qitem{ \Qq{6. Did any of the following objects appear in the video? Choose all that apply.}
\begin{Qlist}
\item Guns
\item Swords
\item Other Weapons
\item Flags
\item Symbols of the Group
\item None of the Above
\end{Qlist}
} \\[-2ex]

\Qitem{ \Qq{7. How likely do you think it is that the people in the video will become involved in the following actions?}
\newline
\begin{tabular}{cccccc}
\textbf{} & Not at All Likely & Not Much Likely & Undecided & Somewhat Likely & Very Much Likely \\
Protests & \Qo & \Qo& \Qo& \Qo & \Qo\\
Violence & \Qo & \Qo& \Qo& \Qo& \Qo\\
Illegal Acts & \Qo & \Qo& \Qo& \Qo& \Qo\\
\end{tabular}
} \\[-2ex]

\end{tabular}
\caption{Video Specific Questions.a}
\label{tab:survey-video-a}
\end{table*}
}

{\renewcommand{\arraystretch}{0.8}
\begin{table*}[t]
\small
\begin{tabular}{p{15cm}}

\Qitem{ \Qq{8. Would you associate the following traits with this video?}
\newline
\begin{tabular}{cccc}
\textbf{} & Yes & Neutral & No \\
Boring (could you pay attention the whole time or not?) & \Qo & \Qo& \Qo\\
Lively (was it energetic? Ex. was there music?) & \Qo & \Qo& \Qo\\
Persuasive (were you convinced by the content?) & \Qo & \Qo& \Qo\\
Trustworthy (did you trust the content?) & \Qo & \Qo& \Qo\\
Logical (was there a structured argument or data presented?) & \Qo & \Qo& \Qo\\
\end{tabular}
} \\

\Qitem{ \Qq{9. Is the video's stance positive, negative, or neutral towards the group?}
\begin{Qlist}
\item Positive
\item Negative
\item Neutral
\end{Qlist}
}

\Qitem{ \Qq{10. Do you think this speaker demonstrated any of the following characteristics?}
\newline
\begin{tabular}{cccc}
\textbf{} & Yes & Neutral & No \\
Charismatic & \Qo & \Qo& \Qo\\
Confident & \Qo & \Qo& \Qo\\
Eloquent & \Qo & \Qo& \Qo\\
Enthusiastic & \Qo & \Qo& \Qo\\
Intelligent & \Qo & \Qo& \Qo\\
Convincing & \Qo & \Qo& \Qo\\
Tough & \Qo & \Qo& \Qo\\
Charming & \Qo & \Qo& \Qo\\
Angry & \Qo & \Qo& \Qo\\
\end{tabular}
} \\

\Qitem{ \Qq{11. Did you enjoy watching the video? }
\newline
\begin{tabular}{ccccccc}
\textbf{} & 1 & 2 & 3 & 4 & 5 & \textbf{} \\
Not at All & \Qo&\Qo &\Qo &\Qo &\Qo & Very Much\\
\end{tabular}
} \\

\Qitem{ \Qq{12. What emotions did you feel when you watched the video? Check all that apply.}
\begin{Qlist}
\item Happiness
\item Sadness
\item Surprise
\item Fear
\item Disgust
\item Anger
\item Confused
\end{Qlist}
} \\

\Qitem{ \Qq{13. Which part of the video was most impactful? (Give the approximate timestamps.) Enter N/A if not applicable.}: \Qline{6cm} } \\

\Qitem{ \Qq{14. Give a short description (a sentence) of the most impactful part of the video you listed above. Enter N/A if not applicable.
}: \Qline{6cm} } \\

\Qitem{ \Qq{15. Do you think any of the content in the video makes a valid point?}
\newline
\begin{tabular}{ccccccc}
\textbf{} & 1 & 2 & 3 & 4 & 5 & \textbf{} \\
Not at All & \Qo&\Qo &\Qo &\Qo &\Qo & Very Much\\
\end{tabular}
} \\

\Qitem{ \Qq{16. Would you take any of the following actions after watching this video? Check all that apply.}
\begin{Qlist}
\item Like the video
\item Dislike the video
\item Post a supporting comment under the video
\item Post a criticizing comment under the video
\item Share the video with friends, families, or on social media platforms
\item Search for similar videos
\item Learn more about the group
\item Consider joining the group
\item Non of the Above
\end{Qlist}
} \\

\end{tabular}
\caption{Video Specific Questions.b}
\label{tab:survey-video-b}
\end{table*}
}

{\renewcommand{\arraystretch}{0.8}
\begin{table*}[t]
\small
\begin{tabular}{p{15cm}}

\Qitem{ \Qq{17. Do you think that others watching this video would consider taking any of the following actions? Check all that apply.}
\begin{Qlist}
\item Like the video
\item Dislike the video
\item Post a supporting comment under the video
\item Post a criticizing comment under the video
\item Share the video with friends, families, or on social media platforms
\item Search for similar videos
\item Learn more about the group
\item Consider joining the group
\item Non of the Above
\end{Qlist}
} \\

\Qitem{ \Qq{18. Did the video change your mind about anything? If so, please elaborate.}:
\newline \Qline{6cm} } \\

\end{tabular}
\caption{Video Specific Questions.c}
\label{tab:survey-video-c}
\end{table*}
}

{\renewcommand{\arraystretch}{0.8}
\begin{table*}[t]
\small
\begin{tabular}{p{15cm}}

\Qitem{ \Qq{1. What is your opinion of the following groups?}
\newline
\begin{tabular}{ccccc}
\textbf{} & Positive & Neutral & Negative & Never Heard of it  \\
QAnon & \Qo& \Qo& \Qo& \Qo \\
Antifa & \Qo& \Qo& \Qo& \Qo\\
Proud Boys & \Qo& \Qo& \Qo& \Qo\\
Oath Keepers & \Qo& \Qo& \Qo& \Qo\\
BLM & \Qo& \Qo& \Qo& \Qo \\
\end{tabular}
} \\

\Qitem{ \Qq{2. Is there anything else about your experience watching these videos that you would like to mention?}: \Qline{6cm} } \\

\Qitem{ \Qq{3. Please rate your experience of this HIT}
\newline
\begin{tabular}{ccccccc}
\textbf{} & 1 & 2 & 3 & 4 & 5 & \textbf{} \\
Much worse than the average HIT & \Qo&\Qo &\Qo &\Qo &\Qo & Much better than the average HIT\\
\end{tabular}
} \\

\Qitem{ \Qq{4. If you would like to give feedback on your experience with this HIT, please do so here.}: \newline \Qline{6cm} } \\

\end{tabular}
\caption{Final Questions}
\label{tab:survey-final}
\end{table*}
}

\twocolumn
\clearpage
\section{Rater Demographics and Background Distribution}
\label{appendix:rater-demographics}

Within the 46 raters participated in the questionnaire:
\begin{itemize}
\item 29 raters were Male, 17 were Female.
\item A major of raters (42) belonged to the 18-29 age group. Only a few (4) belonged to the 30-39 age group.
\item A large number of raters were Asian (37), followed by White (7).
\item 28 raters reported having a Bachelor's degree and 13 raters reported having a Master's degree.
\item About 20 raters reported they were moderates and 19 reported they were liberal.
\item 17 raters agreed slightly to be extroverted and enthusiastic, while others were evenly distributed.
\item 24 raters agreed slightly to be dependable and self-disciplined and no rater strongly disagreed.
\item There was an even distribution of raters who disagreed slightly, neither agreed nor disagreed, agreed slightly to be anxious and easily upset.
\item A major of raters (39) either agreed slightly or strongly to be open to new experiences and complex.
\item There was an even distribution of raters through out all range of disagreement and agreement to be reserved and quiet.
\item 24 raters agreed slightly to be sympathetic and warm and no rater strongly disagreed.
\item 27 raters either disagreed slightly or strongly to be disorganized and careless and no rater strongly agreed.
\item 31 raters either agreed slightly or strongly to be calm and emotionally stable.
\item 23 raters disagreed slightly to be conventional and uncreative, and no rater strongly agreed.
\item 26 raters showed negative opinion on QAnon, 16 raters had never heard of it, and no rater showed positive opinion.
\item 18 raters showed negative opinion on Antifa, 19 raters had never heard of it, and 1 rater showed positive opinion.
\item 23 raters showed negative opinion on Proud Boys, 21 raters had never heard of it, and no rater showed positive opinion.
\item A major of raters (35) had never heard of Oath Keepers, and no rater showed positive opinion.
\item 18 raters showed positive opinion on BLM, 15 raters were neutral, and 3 raters showed negative opinion.
\item 27 raters did not trust Fox News, 14 raters were neutral, and 1 rater trusted it.
\item 28 raters had never heard of Breitbart News and 11 raters did not trust it.
\item 21 raters were neutral on MSNBC News and 11 raters trusted it.
\item 28 raters either trusted or were neutral on PBS News and 3 raters did not trust it.
\item 29 raters either trusted or were neutral on Associate Press News and 1 raters did not trust it.
\item 29 raters either trusted or were neutral on NPR and one raters did not trust it.
\item A major raters (44) either trusted or were neutral on The Wall Street Journal and 2 raters did not trust it.
\item A major raters (39) either trusted or were neutral on CNN and 7 raters did not trust it.
\end{itemize}

\begin{figure*}[!htb]
\includegraphics[width=.3\textwidth]{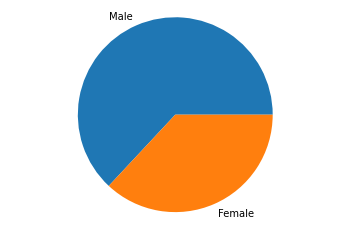}\hfill
\includegraphics[width=.3\textwidth]{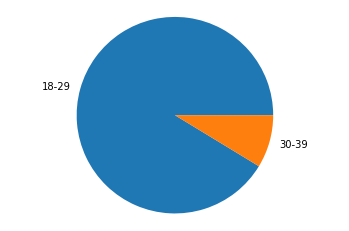}\hfill
\includegraphics[width=.35\textwidth]{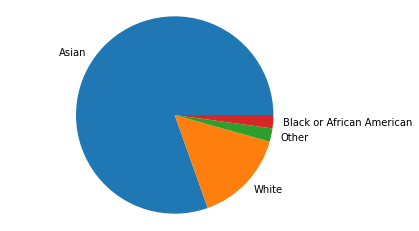}\\
\centering
\includegraphics[width=.33\textwidth]{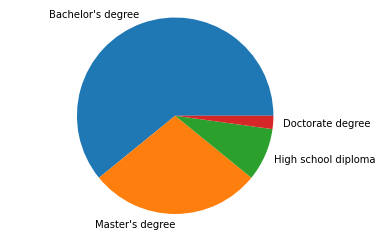}
\includegraphics[width=.31\textwidth]{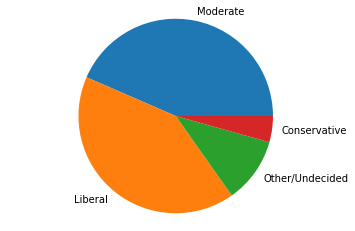}
\caption{Rater demographics. A total of 46 raters completed the questionnaire.}
\label{fig:demographics}
\end{figure*}

\begin{figure*}[!htb]
\centering
\includegraphics[width=1.\textwidth]{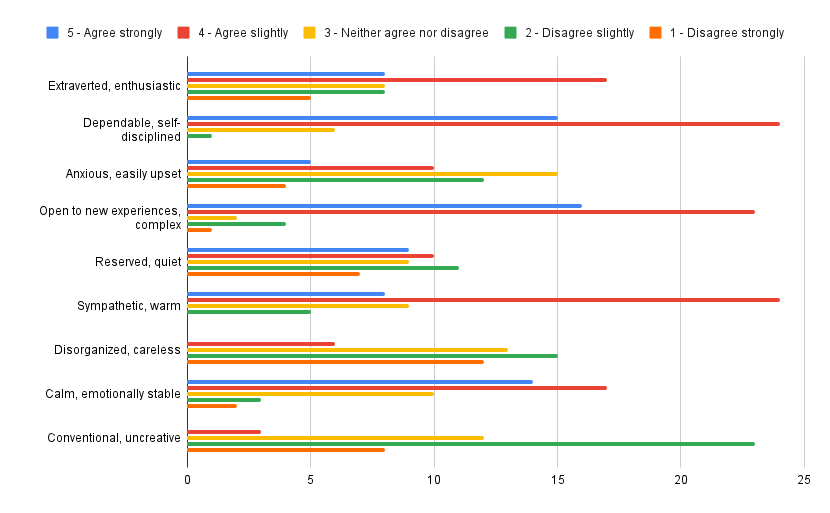}\hfill
\caption{Rater self-reported personalities. A total of 46 raters completed the questionnaire.}
\label{fig:self-report-personality}
\end{figure*}

\end{document}